\documentclass[journal,twoside,web]{ieeecolor}
\usepackage{tmi}
\usepackage{cite}
\usepackage{float}
\usepackage{amsmath,amssymb,amsfonts}
\usepackage{multirow}
\usepackage{booktabs}
\usepackage{makecell}
\usepackage{algorithmic}
\usepackage{graphicx}
\usepackage{textcomp}
\usepackage{graphicx}
\usepackage{subfigure}

\newcommand{\BSpace}{\vspace{-.05in}}
\usepackage{overpic}
\usepackage{rotating,soul}
\usepackage{gensymb}
\newcommand{\figref}[1]{Fig.~\ref{#1}}
\newcommand{\tabref}[1]{Tabel~\ref{#1}}
\newcommand{\equref}[1]{Eq.~(\ref{#1})}

\def\BibTeX{{\rm B\kern-.05em{\sc i\kern-.025em b}\kern-.08em
    T\kern-.1667em\lower.7ex\hbox{E}\kern-.125emX}}

\begin{document}
\title{Unsupervised Domain Adaptation for Retinal Vessel Segmentation with Adversarial Learning and Transfer Normalization}
\author{
  Wei Feng, 
  Lie Ju,
  Lin Wang, 
  Kaimin Song,
  Xin Wang,
  Xin Zhao, 
  Qingyi Tao,
  and Zongyuan Ge

\thanks{(Corresponding author: Zongyuan Ge)}
\thanks{Wei Feng, Lin Wang, Kaimin Song are with the Monash-Airdoc joint research group, Monash University, Clayton, VIC 3800 Australia (E-mail: {wf02429,wanglin.mailbox,kims75699}@gmail.com).}
\thanks{Lie Ju and Zongyuan Ge are with Monash University, Clayton, VIC 3800 Australia. Lie Ju and Zongyuan Ge are also with Airdoc, Beijing 100089, China (E-mail: julie334600@gmail.com, zongyuan.ge@monash.edu).}
\thanks{Qingyi Tao is with NVIDIA AI Technology Center, Singapore (E-mail: qtao002@e.ntu.edu.sg).}
\thanks{Xin Zhao and Xin Wang are with Airdoc, China (E-mail: {zhaoxin,wangxin}@airdoc.com).}
}
\maketitle
\begin{abstract}
Retinal vessel segmentation plays a key role in computer-aided
screening, diagnosis, and treatment of various cardiovascular and ophthalmic diseases.
Recently, deep learning-based retinal vessel segmentation algorithms have achieved
remarkable performance.
However, due to the domain shift problem, the performance of these algorithms often
degrades when they are applied to new data that is different from the training data.
Manually labeling new data for each test domain is often a time-consuming and laborious
task. In this work, we explore unsupervised domain adaptation in retinal vessel
segmentation by using entropy-based adversarial learning and transfer normalization layer
to train a segmentation network, which generalizes well across domains and requires no
annotation of the target domain. Specifically, first, an entropy-based adversarial
learning strategy is developed to reduce the distribution discrepancy
between the source and target domains while also achieving the
objective of entropy minimization on the target domain. In addition, a new transfer
normalization layer is proposed to further boost the
transferability of the deep network. It normalizes the features of
each domain separately to compensate for the domain distribution gap.
Besides, it also adaptively selects those feature channels that are
more transferable between domains, thus further enhancing the
generalization performance of the network.  We conducted extensive experiments on three
regular fundus image datasets and an ultra-widefield fundus image dataset,
and the results show that our approach yields significant performance
gains compared to other state-of-the-art methods.
\end{abstract}

\begin{IEEEkeywords}
Retinal vessel segmentation, Domain adaptation, Batch normalization, Ultra-wide field fundus images.
\end{IEEEkeywords}

\section{Introduction}
\label{sec:introduction}
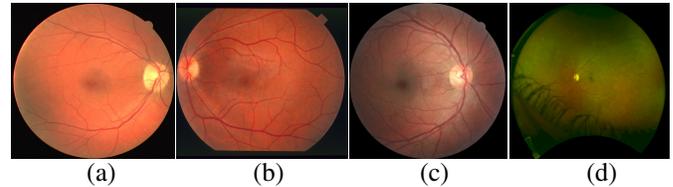
\begin{figure}[htbp!]
    \begin{overpic}[width=\linewidth]{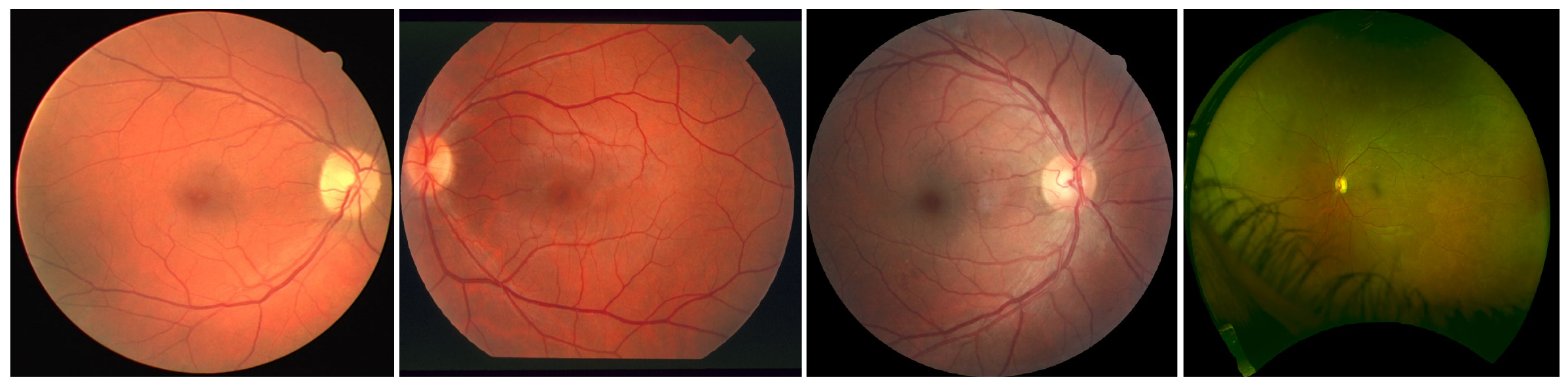}
        \put(12,-3){(a)}
        \put(37,-3){(b)}
        \put(62,-3){(c)}
        \put(87,-3){(d)}
    \end{overpic} \\
    \vspace{-.10in}
    \caption{Exemplar fundus images from different fundus datasets. From left to right are (a)DRIVE; (b)STARE; (c)our clinical dataset; (d)ultra-wide field fundus image dataset. It can be observed that there is a distinct visual appearance between different fundus datasets.
    }
    \label{fig.1}
\end{figure}
\IEEEPARstart{V}{essel} analysis in retinal fundus images is one of the most important tools for
screening, diagnosis, and treatment of many ocular, hypertension, and cardiovascular diseases
\cite{srinidhi2017recent}. For example, diabetic retinopathy can cause neovascularization,
hypertension can cause vessel curvature, and stroke and atherosclerosis can cause structural changes
in vessel morphology \cite{patton2005retinal}. Accurate retinal vessel segmentation is crucial for
subsequent vessel morphology analysis \cite{jin2019dunet}. Recently, deep learning-based approaches
have made significant progress for automated retinal vessel segmentation tasks
\cite{lian2019global,wu2018multiscale}. However, a high-performance deep learning model usually relies
on a large number of training samples with corresponding annotations. This is impractical in the field of medical
imaging, especially for retinal vessel segmentation. Even for experienced clinicians, manually
annotating a binary vessel map is very time-consuming and labor-intensive. For example, it
typically takes 18 hours to fully annotate an ultra-widefield (UWF) fundus image. Therefore, it is of importance to explore the use of fundus images from other modalities for scalable and robust modeling
\cite{ding2020weakly}.

However, fundus images from different sources have distinct visual appearance.
As shown in \figref{fig.1}, these differences in visual appearance are due to
differences in imaging devices and imaging protocols. This is common in
practical clinical scenarios. Even fundus images obtained from the same fundus
camera may show variation because of the differences in clinical settings and
subjects. In addition, the range of imaging and resolution of these fundus
images from different sources may also differ. For example, the regular
fundus images in \figref{fig.1}(a)--(c) have an imaging angle of only
20$\degree$--60$\degree$ and can only cover a small part of the central retinal
area, which prevents physicians from adequately identifying many common fundus
diseases due to the lack of peripheral fundus angiography imaging. In contrast,
the UWF fundus image in \figref{fig.1}(d) can be obtained at
200$\degree$, allowing imaging of approximately 80\% of the retinal fundus. In
addition, it can show both posterior pole and peripheral retinal images without dilating the
patient's pupil, which can reveal occult peripheral retinal ischemia or neovascularisation, thus
more effectively guiding relevant treatment plans and expanding clinicians' knowledge of the peripheral retina \cite{ju2021leveraging}. 
However, these differences between fundus datasets  will lead to substantial performance degradation when transferring a deep learning model trained on one fundus dataset to another, since a high-performance deep learning model requires that the train data and test data are drawn from the same distribution \cite{pan2009survey}.

To address these challenges mentioned above, this paper proposes a new unsupervised domain adaptation framework with adversarial learning and transfer normalization for cross-domain retinal vessel segmentation. Specifically, We first propose an entropy-based
adversarial learning strategy to align the entropy of the predicted outputs from the source and
target domains, which has two benefits: (1) reducing the distribution gap between the source and target domains, thus enabling
the model trained with the source domain data to generalize
well on the unlabeled target domain; (2) achieving the goal of
minimizing the entropy on the target domain, forcing the model to
produce confident predictions for unlabeled data from the target
domain. Our motivation stems from the observation that models trained on labeled source domain
data tend to produce high confidence predictions on source domain data and low confidence 
predictions on target domain data due to domain distribution differences. Here, we employ
Shannon Entropy \cite{shannon1948mathematical} to model prediction uncertainty, with a high
confidence prediction implies a low entropy value and a low confidence prediction implies a
high entropy value. We argue that it would be a possible solution to compensate for domain distribution differences to use adversarial learning by aligning entropy maps between domains,
thereby encouraging models to produce high confidence (low entropy) predictions on unlabeled
target domain data as well. 

In addition, to further improve the network
transferability, a transfer normalization layer is proposed to
replace the batch normalization (BN) layer in deep neural network
(DNN) for domain adaptation setting. The transfer normalization layer normalizes feature representations from different domains separately, which
can compensate for the discrepancies between domains. At the same time, since different feature channels capture different patterns, they have different transferability, and thus need careful modulated during the adaptation process. To this
end, we use the statistics obtained from the BN layer to measure the transferability of
different feature channels between the source and
target domains, and then adaptively reweight each feature channel so that feature channels with
higher transferability can be emphasized, and those with lower transferability are suppressed, which enables further enhancing the generalization performance of the model on the target domain.
The main contributions of this paper are as follows: 
\begin{itemize}

\item [1)] We propose a new unsupervised domain adaptation framework for
cross-domain retinal vessel segmentation. An entropy-based adversarial training strategy is developed to reduce the domain gap between the source and target domains, thus
improving the generalization performance of the model on the target domain.
Besides, it achieves the goal
of entropy minimization and thus encourages the model to produce high
confidence predictions on the target domain.
  \item [2)]
To further enhance the network's transferability, we propose a transfer normalization layer which
normalizes the feature representations of two domains separately to explicitly compensates for the inter-domain differences. We also quantified the transferability of each feature channel and adaptively reweighted them, thus a  further improvement on the generalization performance of the network was observed.
  \item [3)] To the best of our knowledge,
this is the first work that explores the use of regular fundus datasets to
aid in UWF retinal vessel segmentation. Our method can detect most of the
retinal vessels in UWF fundus images without any annotation on the target dataset.
  \item [4)]
We evaluated our method on three regular fundus datasets and a UWF fundus dataset.
The experimental results show that our method not only yields significant performance gains
on transfer tasks between regular fundus datasets, but also outperforms other
state-of-the-art methods by a large margin on the more challenging transfer tasks between regular and UWF fundus datasets.
\end{itemize}

\section{RELATED WORK}
\label{sec:related work}
\subsection{Retinal vessel segmentation}
The existing retinal vessel segmentation can be divided into two main branches: unsupervised methods
and supervised methods. Unsupervised methods use heuristics knowledge about retinal vessels to
identify blood vessels, mainly including the shape of the vessels, the position of the optic cup and optic
disc, and the centerline of the vessels, etc. Methods derived from this heuristic knowledge include
matched filter \cite{al2007improved}, vessel tracking \cite{wink2004multiscale},
morphological processing \cite{fraz2011retinal}, etc. However, the acquisition of such
heuristic knowledge is very challenging and subjective, and the generalization performance of these
methods is also very poor in practical clinical application scenarios \cite{moccia2018blood}.
Supervised learning methods mainly include traditional machine learning methods and deep learning
methods. Traditional machine learning methods use the labeled retinal vessel image dataset and hand-crafted features to build segmentation models. For example, Soares et al. \cite{soares2006retinal}
extracted hand-crafted features using a 2-D Gabor filter and then used a Bayesian network
to classify each pixel in the image. 
% osareh et al. \cite{osareh2009automatic} extracted features using
% a multiscale filter, then used principal component analysis for dimensionality reduction, and finally
% used a hybrid model of Gaussian mixture model (GMM) and support vector machine (SVM) for
% classification.
Roychowdhury et al. \cite{roychowdhury2014blood} extracted digital morphological
features of vessel regions and background regions and then used Gaussian mixture model (GMM) for classification. However, the
performance of these methods relies heavily on the quality of the manually extracted features. Deep
learning methods are capable of automatically extracting features from the original fundus image and
are able to perform vessel segmentation in an end-to-end manner. For example, Yan et al.
\cite{yan2018joint} proposed a new segmentation loss to balance the importance between thick and thin
vessels during training and thus better segment fine vessels.
% Li et al. \cite{li2020joint} proposed a
% novel segmentation network called SeqNet, which first distinguishes blood vessels from the background
% and then classifies them into arteries and veins. A post-processing technique was also proposed to
% further refine the segmentation result.
Son et al. \cite{son2017retinal} introduced a generative
adversarial network to produce more accurate and clearer vessel maps, and the experimental results
demonstrated that their method could generate fewer false predictions on fine vessels. 
Sun et al. \cite{sun2020robust} developed a  channel-wise stochastic gamma correction and a channel-wise
stochastic vessel augmentation technique to encourage the model to learn invariant and more
discriminative features. However, all of the above methods require a large amount of annotated
training data to train a high-performance vessel segmentation model, and the trained models are
difficult to generalize across different fundus datasets due to the domain shift.  In this work, we
focus on unsupervised domain adaptation in retinal fundus vessel segmentation, leveraging annotated source fundus
datasets to build a  robust segmentation model that generalizes well over unannotated target fundus datasets with
domain shift.

\subsection{Domain adaptation}
Domain adaptation aims to bridge the distribution gap between the source and target domains so that
the model can generalize well across domains. In this paper, we focus on deep learning-based domain
adaptation methods. Previous deep domain adaptation methods generally fall into two main categories:
moment matching and adversarial training \cite{peng2019moment}. Moment matching aims at aligning the
feature distributions of the source and target domains by minimizing the difference between the
distributions of the two domain data in the high-dimensional feature space
\cite{long2018transferable,peng2019moment}. Adversarial training is derived from the idea of generative
adversarial network (GAN), where generator and discriminator boost each other by establishing a
two-player min-max competition game mechanism \cite{goodfellow2014generative,tsai2018learning}. Domain
adaption has achieved impressive performance in several areas such as computer vision and natural
language processing \cite{tsai2018learning}. However, in the field of retinal vessel
segmentation, domain adaption has rarely been studied. Mehran et al. \cite{javanmardi2018domain}
proposed a domain adaptation network based on adversarial learning for cross-domain retinal vessel
segmentation. Zhuang et al. \cite{zhuang2019domain} proposed an asymmetrical maximum classifier
discrepancy (AMCD) approach for cross-domain retinal vessel segmentation.

However, these methods only evaluated the performance on the transfer tasks between limited regular
fundus datasets and did not consider transfer tasks between regular fundus datasets and UWF fundus
datasets. UWF images have more valuable clinical applications than regular fundus images \cite{ju2021leveraging}. However,
models trained with regular fundus images may not be able to accurately identify peripheral vessels
and some lesions in UWF fundus images due to the difference in imaging range. In addition, UWF images
contain more noise interference such as eyelashes, eyelids, and lesions, which leads to the much
higher cost of labeling UWF images. Therefore, it is of interest to investigate how to utilize regular
fundus image datasets with rich annotation to assist in the annotation of UWF fundus datasets. To our
best knowledge, we are the first attempt to utilize regular fundus images to perform UWF vessel
segmentation without any annotations of UWF fundus images. Our
approach was able to identify most retinal vessels of the UWF fundus images and achieved consistent
performance gains in all evaluation metrics.

\subsection{Batch normalization}
Batch normalization (BN) \cite{ioffe2015batch} is one of the most successful inventions in deep neural
networks. It can accelerate network training and alleviate the problem of internal covariance shift
during the training process \cite{santurkar2018does}. However, the traditional BN layer may not be
suitable for the domain adaptation problem because it combines the source and target domain data into
a single batch and feeds the combined data to the BN layer during training so that the statistics (mean and variance)
of the BN layer are estimated on the data of both domains without considering which domain the feature
representation comes from. However, due to domain shift \cite{quionero2009dataset}, the distribution
of source and target domain are different, which means that their distribution statistics (mean and variance) are also different. In this case, the traditional BN layer may distort the feature
distribution and worsen the network's transferability. Some research work has been done to address
this issue. For example, Li et al. \cite{li2016revisiting} proposed a simple yet effective strategy for domain adaptation problem by replacing the statistics of each BN layer in the source domain with the
statistics of the target domain. Liu et al. \cite{liu2020ms} proposed a domain-specific batch
normalization layer by normalizing the features of each domain separately during the training process for improving the performance of prostate
segmentation. However, these methods do not explore
the transferability among each channel of different domains. Since some feature channels learn
domain shared patterns that should be emphasized, while some feature channels may learn non-shared
patterns, such as background or noise, etc., and thus should be suppressed. In this paper, we measure
the similarity of different feature channels across domains by the mean and variance estimated in the BN
layer so that each feature channel can be reweighted adaptively. Feature channels with higher domain relevance will be highlighted, while those with lower domain relevance will be repressed, thus further enhancing the transferability of the network.

\begin{figure*}[htbp]
    \centering
    \includegraphics[width=\linewidth]{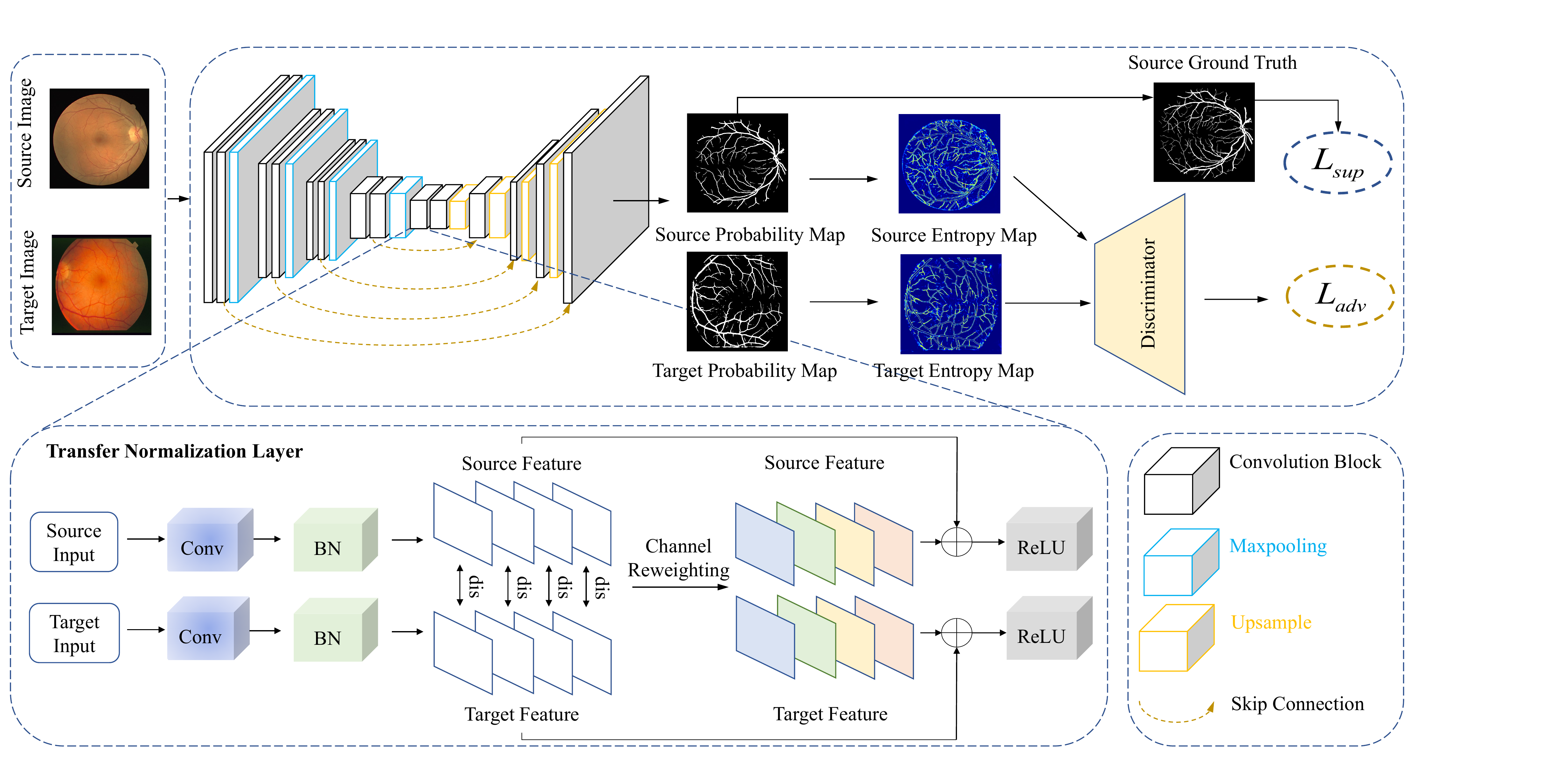}
    \vspace{-.15in}
    \caption{The overview of our proposed framework for unsupervised domain adaptation in retinal
    vessel segmentation. It uses an entropy-based adversarial learning strategy to bridge the domain
    distribution discrepancy while prompting the model to produce confident predictions on the
    unlabeled target domain data. In addition, a new transfer normalization layer is proposed to
    normalize the feature representation of each domain separately while adaptively reweighting each
    feature channel to further enhance the generalization performance of the model.}
    \label{fig.2}
\end{figure*}
\section{METHODOLOGY}
 \figref{fig.2} demonstrates the overview of our proposed framework for unsupervised domain adaptation in retinal vessel segmentation, which performs domain adaptation between two fundus image datasets from different sources by an
entropy-based adversarial training method and a transfer normalization layer. The entropy-based
adversarial training strategy can reduce domain differences while adjusting the model to produce
confident predictions on unlabeled target fundus images. The transfer normalization layer can further
improve the generalization performance of the network. 
\subsection{Entropy-based Adversarial Learning}
In this work, we
follow the principle of adversarial training and propose an entropy-based adversarial learning
strategy to train our domain adaptation framework. The spirit of adversarial learning stems from generative adversarial networks
\cite{goodfellow2014generative}, which consists of two modules: a generator and a discriminator, where
the generator learns to capture the distribution of real data while the discriminator tries to
distinguish the generated data from the real data. The two networks are alternately optimized until the data generated by the generator can fool the discriminator. 

Specifically, given the labeled source domain
fundus dataset $D^s=\{(x_i^s, y_i^s)\}_{i=1}^{N_s}$ and the unlabeled target domain fundus dataset
$D^t=\{x_i^t\}_{i=1}^{N_t}$,where $x_i^{s}\in \mathbb{R}^{H \times W\times 3}$ and $x_i^{t}\in \mathbb{R}^{H \times W\times 3}$ denote the $i$-th
sample of the source domain and target domain, and $y_i^s\in \{0,1\}^{H \times W}$ is the corresponding
ground truth label. 
We feed the source and target domain data into a segmentation network $F$ and obtain the predicted
probability maps $P^s\in \mathbb{R}^{H \times W  \times C}$ and $P^t\in \mathbb{R}^{H \times W \times
C}$ for each domain, $C=2$ is the number of categories. We then generate  the entropy map $I^s$ and
$I^t$ based on the probability maps for each domain: 
\begin{equation}\label{equ:ent}
\begin{aligned}
I^{s}=-\frac{1}{\log(C)} \sum_{c \in \{0,1\}}P_{c}^{s}\log{P_{c}^{s}}, \\
I^{t}=-\frac{1}{\log(C)} \sum_{c \in \{0,1\}}P_{c}^{t}\log{P_{c}^{t}}.
\end{aligned}
\end{equation}

A discriminator network $D$ is then constructed to distinguish the entropy maps from the two domains.
 Unlike previous work \cite{tsai2018learning}, we use the entropy map rather
 than the original prediction probability map as the input to the discriminator.
 This is because models trained on labeled source domain data tend to give
 over-confident (low-entropy) predictions for source domain data and
 under-confident (high-entropy) predictions for unlabeled target domain data.
 Therefore, we believe that by aligning the entropy map between domains with adversarial training can encourage the model to give confident predictions on
 unlabeled target domain data and reduce the distribution differences between
 domains so that the segmentation model generalizes well on the target domain.
%  The segmentation network and the discriminator compete with each other until the segmentation network
% learns a domain invariant feature that fools the discriminator.
The discriminator uses a binary cross-entropy loss to classify the entropy maps of the source and target domains, which can be
formulated as:
\begin{equation}\label{equ:discriminator}
L_d=-\sum_{k\in \mathcal{V}}(1-z) \log (D(I_{k}^{s})) +z \log (D(I_{k}^{t})),
\end{equation}
where $\mathcal{V}=\{1,2, \ldots, H \times W\}$ denotes all pixels in the fundus image, and $z=0$ if the entropy map comes from the source domain and $z=1$ if the entropy map comes from the target domain.

For the segmentation network, we expect the distribution of $I^t$ to be close to $I^s$ .The optimization objective of the segmentation network can be formulated as: 
\begin{equation}\label{equ:segmentation}
\begin{aligned}
L_{seg} & = L_{sup} + \lambda_{d} L_{a d v}\\
 & = -\sum_{k \in \mathcal{V}}\sum_{c \in \{0,1\}} P_{k,c}^{s}\log ( y_{k,c}^{s}) -\lambda_{d} \sum_{k\in \mathcal{V}} \log ({D}(I_{k}^{t})),
\end{aligned}
\end{equation}
where $L_{sup}$ is the supervised loss on the labeled source domain samples and $ L_{a d v}$ is the adversarial loss. $\lambda_{d}$ is a hyperparameter to balance these two loss terms, which is fixed as 1$e-$3 in our setting.
\subsection{Transfer Normalization}
Batch normalization (BN) \cite{ioffe2015batch} is widely used in deep neural networks to accelerate
network convergence and reduce internal covariance shift. Recent studies have shown that the essence
of BN works due to its ability to make the loss landscape smoother \cite{santurkar2018does}. 
Specifically, let $h_m \in [h_1,h_2,\cdots,h_M]$ denote one of the $M$ feature channels at a certain 
layer of the deep network, BN first calculates the mean and variance of $h_m$ over a mini batch of 
data, i.e. $E[h_{m}]$ and $\operatorname{Var}[h_{m}]$, each feature channel is then transformed to 
zero mean and unit variance using this mean and variance. To enhance the representation capability of 
the model, two additional learnable parameters $\gamma$ and $\beta$ are introduced to scale 
and shift each normalized feature channel:
\begin{equation}\label{equ:bn_t}
o_{m} = \gamma_{m} \widehat{h}_{m}+\beta_{m}, ~~~\widehat{h}_{m} = \frac{h_{m}-E[h_{m}]}{\sqrt{\operatorname{Var}[h_{m}]+\epsilon}},
\end{equation}
where $\epsilon$ is an  infinitesimal to ensure numerical stability.

However, in our problem setting, the imaging devices, imaging protocols, range of imaging, and resolution vary across different fundus datasets, resulting in significant differences between the
feature distributions of different fundus datasets, which is often also referred to as the
domain shift phenomenon \cite{quionero2009dataset}. The traditional BN layer combines data from
different domains into a single batch and then normalizes this single batch, which may lead to two
issues: (1)due to domain distribution gap, the sharing of mean and variance among different domains in the BN layer may distort the feature distribution and deteriorate the transferability of the network, making it difficult for the network to learn generic and robust features; (2) during the training phase, the BN layer estimates global mean and variance on the single batch consisting of data from different domains, which may be inaccurate and may cause performance degradation when used directly for normalization of target domain data during the testing phase.

To address the above issues, we propose a transfer normalization layer to replace the BN layer for the
domain adaptation problem. Specifically, let $h_m^{s} \in
[h_1^{s},h_2^{s},\cdots,h_M^{s}]$ and $h_m^{t} \in
[h_1^{t},h_2^{t},\cdots,h_M^{t}]$ denote a certain feature channel of the feature map from the
source and target domain, and we normalize the feature channel for each domain separately as follows:
\begin{equation}\label{equ:tn_1}
\begin{aligned}
 o_{m}^{s} = \gamma_{m}^{s} \widehat{h}_{m}^{s}+\beta_{m}^{s},
 ~~~\widehat{h}_{m}^{s} = \frac{h_{m}^{s}-E[h_{m}^{s}]}{\sqrt{\operatorname{Var}[h_{m}^{s}]+\epsilon}},\\
 o_{m}^{t} = \gamma_{m}^{t} \widehat{h}_{m}^{t}+\beta_{m}^{t},
 ~~~\widehat{h}_{m}^{t} = \frac{h_{m}^{t}-E[h_{m}^{t}]}{\sqrt{\operatorname{Var}[h_{m}^{t}]+\epsilon}},
\end{aligned}
\end{equation}
where $[\gamma^{s},\beta^{s}]$ and $[\gamma^{t},\beta^{t}]$ are the learnable parameters of the source and target domains, respectively.

Meanwhile, for the domain adaptation problem, since the patterns captured by different feature
channels are different, some feature channels may capture patterns shared among both domains, while others
capture domain-specific patterns. Therefore, a heuristic question is: Can we quantify the
transferability of different feature channels and thus enhance the transferability of the network in
some way? In this paper, we provide a solution by reusing the statistics computed by the BN layer on
the feature representation of each domain to automatically select those feature channels that are more
transferable. Specifically, we first compute the distance between each channel of the source and
target domains as follows:
\begin{equation}\label{equ:tn_2}
d_{m}=|\frac{E[h_{m}^{s}]}{\sqrt{\operatorname{Var}[h_{m}^{s}]+\epsilon}}-\frac{E[h_{m}^{t}]}{\sqrt{\operatorname{Var}[h_{m}^{t}]}+\epsilon}|,
\end{equation}
$d$ quantifies the transferability of each feature channel, then we use Student t-distribution \cite{student1908probable} with one degree of freedom to convert the computed distance into probability $\eta$ to adaptively reweight each feature channel: 
\begin{equation}\label{equ:tn_3}
\eta_{m}=\frac{M(1+d_{m})^{-1}}{\sum_{j=1}^{M}(1+d_{j})^{-1}},
% , \quad k=1,2, \ldots, K
\end{equation}
where $M$ is the number of feature channels in a layer of the network. Moreover, to avoid over-penalizing those
informative channels, we further introduce a residual
mechanism \cite{he2016deep} to combine the normalized and reweighted feature channels to form the final output:
\begin{equation}\label{equ:tn_4}
\widehat{o}^{s}=(1+\eta) o^{s}, \quad \widehat{o}^{t}=(1+\eta) o^{t}.
\end{equation}

Using this distance-based probability to quantify the
transferability of each feature channel, those channels with
higher transferability will be assigned with greater
importance, while those with lower transferability will be
assigned with lower weights. This mechanism allows the network to better learn domain shared features while
suppressing the interference of those non-essential domain-specific variations.
\section{EXPERIMENTS}
\label{sec:experiments}
To validate the effectiveness of our proposed method, we
perform experiments on two types of transfer tasks: (1)
transfer tasks between several regular fundus datasets;
(2) transfer tasks between regular fundus datasets and
UWF fundus dataset, which is a more challenging transfer task.
\subsection{Datasets}
We conducted experiments on three regular fundus datasets (DRIVE, STARE, a regular fundus dataset  obtained from a private hospital) and a UWF fundus dataset (PRIME-FP20).  The details of each fundus dataset are as follows.

The DRIVE (digital retinal images for vessel extraction) dataset\cite{staal2004ridge} is a database of
retinal images constructed by the Diabetic Retinopathy Screening Group in the Netherlands, which was
derived from 400 diabetic patients aged 20-90 years. It contains a total of 40 retinal images, 33 of
which have no diabetic retinopathy, and the remaining 7 have diabetic retinopathy. All images were
acquired with a CannonCR5 astigmatism-free 3CCD camera with an acquisition angle of 45\degree ~and a
resolution of $584\times565$. The dataset contains a training subset and a test subset, each with 20
images. Each training image is equipped with a manually annotated binary vessel map as the ground
truth label, and each test image is equipped with two manually annotated binary vessel maps, one for
the ground truth label and one for the human observer's results.

The STARE (structured analysis of the retinal) dataset\cite{hoover2000locating} contains 20 retinal fundus images, 10 of which have no retinal fundus lesions, and the remaining 10 have retinal fundus lesions. Each image was taken with the TopCon TRV-5 fundus camera at an angle of 35\degree ~and a resolution of $700\times605$. Each image has two manually labeled binary vessel maps by human experts, where the first one is usually used as the ground truth label. Since there is no official split for this dataset, we use the first ten images for training and the remaining ten images for testing.

Retinal40 is a regular fundus dataset obtained from a private hospital. It contains 40 regular fundus images, 30 of which have no retinal fundus lesions and 10 of which have retinal fundus lesions. Each fundus image was acquired with a 45\degree  ~fundus camera (Type CR6-45NM, Canon Inc, Lake Success,
NY, United States). We resize each image to $1024 \times 1024$ pixels. Each fundus image was annotated by a human annotator using ImageJ software \cite{schindelin2012fiji} with the segmentation editor plugins to annotate the ground truth labels. The annotator repeatedly adjusted the image illumination and contrast to accurately label the retinal vessels in the fundus images. We also provide a binary mask for the FOV of each fundus image. We use the first 30 images for training and the remaining 10 images for testing.

The PRIME-FP20 dataset\cite{ding2020prime} contains 15 pairs of simultaneously captured UWF-FP and UWF-FA images. In this paper, we only use UWF-FP images. All images were captured using the Optos California and 200Tx cameras (Optos plc, Dunfermline, United Kingdom). Each image has a resolution of $4000\times4000$ pixels and is stored in 8-bit TIFF format. Each UWF FP was equipped with a binary vessel map labeled by a human expert and a binary mask to distinguish fundus regions from non-fundus regions. We use the first ten images for training and the remaining five images for testing.

\subsection{Pre-processing and Implementation Details}
Since different fundus datasets have different illumination,
contrast, and resolution, to compensate for this variation, we first convert each image to grayscale, then normalize it to zero mean and unit variance, and finally map it to $[0, 255]$. We also use Contrast
Limited Adaptive Histogram Equalization (CLAHE) \cite{setiawan2013color} to further improve
the contrast and suppress noise in the fundus images.
Specifically, we divide each fundus image into several
$8\times8$ regions and then perform histogram equalization
on each small region individually. Compared with the
traditional histogram equalization method, CLAHE is more
suitable for improving the local contrast and enhancing the
image edge information. In addition, following our previous work \cite{ju2021leveraging}, we removed artifacts in the UWF fundus images, such as eyelids and eyelashes, using a UNet \cite{ronneberger2015u} segmentation network.

We construct two types of transfer tasks, i.e., transfer tasks between the regular fundus dataset and transfer tasks
between the regular fundus and UWF fundus datasets.
For each transfer task, we train the segmentation model
using all the labeled source-domain fundus images and the
unlabeled target-domain training images, and then evaluate
the performance of the model using the testset of the
target domain, noting that we only use the testset of target-domain
 in the testing phase. For example, for the
transfer task $DRIVE \to  STARE$, we train the
segmentation model using 40 labeled fundus images of 
DRIVE(source domain) and ten unlabeled training images
of STARE(target domain), and then evaluate the model performance on the testset of STARE .

For model training, following previous work \cite{wu2018multiscale}, we use the patch-based training approach. Specifically, we randomly extract 10,000 $64\times64$ patches from each fundus image in each fundus dataset for model training, so
the number of patches of each dataset are: DRIVE (400,000), STARE (200,000), Retinal40 (400,000), and UWF
(150,000), respectively. In the test phase, we also use a patch-based evaluation strategy. Specifically, we extract partially overlapping patches from each test image in both directions with a
stride of 10 in an orderly manner, and each patch has a size of $64\times64$. We then feed the extracted
patches into the segmentation network to obtain the prediction probability map. Since the patches are
partially overlapping, some pixels may appear in multiple patches, so we further average the
prediction probability of each pixel to obtain the final binary vessel prediction map.
% The patched-based training and evaluation strategy has two benefits: (1) it increases the number of
% training samples, which acts as a role of data augmentation and alleviates the overfitting phenomenon.
% Since these fundus datasets all have only a limited number of fundus images, feeding them directly
% into the model for training would lead to rapid overfitting of the model; (2) The details and useful
% information in fundus images can be maximally preserved. Since fundus images from different domains
% have different resolutions, e.g., regular fundus images have lower resolution while UWF fundus images
% have higher resolution, previous work \cite{javanmardi2018domain,zhuang2019domain} directly resize
% fundus images from different domains into the same size, which leads to a significant loss of useful
% information and we can avoid this problem by using this patch-based training method.
During the training phase, to alleviate the overfitting phenomenon and further improve the generalization
performance of the model, we also use some data augmentation techniques, including: random
horizontal/vertical flipping, random rotation at an angle from $[90\degree,
180\degree, 270\degree]$\cite{wu2018multiscale}. 

We adopted UNet \cite{ronneberger2015u} as the segmentation network for its
superior performance in several medical image processing
tasks. Unlike the original UNet, we replace the BN layer following each convolutional layer with
our proposed transfer normalization layer for the domain
adaptation problem. In addition, for the discriminator
network, we use a structure similar to DCGAN \cite{radford2015unsupervised}. Specifically, it contains four convolutional layers, each followed by a
leaky ReLU activation layer, and finally the discriminator
outputs the probability that the input belongs to the source and target domains.
We use the Adam algorithm \cite{kingma2014adam} to optimize the segmentation network and discriminator
network, with the learning rate set to 1$e-$3 and the weight decay set to 3$e-$4 for the segmentation
network, and 1$e-$4 for the learning rate of the discriminator network. The batch size is 256, and the
number of iterations is 50,000. To schedule the learning rate, we apply a polynomial decay to the
learning rate of the segmentation network and the discriminator \cite{chen2017deeplab}. All models are
implemented based on the PyTorch deep learning framework \cite{paszke2017automatic}, and are trained
and evaluated on four 1080Ti GPUs. Following \cite{wu2018multiscale}, we adopt six metrics to evaluate the performance of our proposed approach, including:the area under the receiver operating characteristics curve
(AUC), area under the precision-recall curve (AUPR), F1 score (F1), sensitivity (SE), specificity (SP), and accuracy (ACC) to evaluate the performance of different algorithms. 

\subsection{Results}
% \subsubsection{Segmentation Performance of Transfer Tasks between regular fundus Datasets} \Lie{Sub-title is too long}\label{sec:seg_regular}
\subsubsection{Evaluations on the Transfer Tasks Regular$\to$ Regular}\label{sec:seg_regular}
We compare our proposed algorithm with the baseline model, i.e., model trained with source domain data only and then tested on unlabeled target domain data, and five state-of-the-art deep domain adaptation methods,
including: DANN \cite{javanmardi2018domain}, AdaptSeg \cite{tsai2018learning}, AdaBN \cite{li2016revisiting}, AMCD \cite{zhuang2019domain}, and DSBN \cite{liu2020ms}. DANN introduces a discriminator network to distinguish between source and target domain feature representations, while the segmentation network learns a domain invariant feature to fool the discriminator. AdaptSeg employs adversarial learning in the output space to bridge domain distribution gaps. AdaBN substitutes the statistics of the BN layer with those of the target domain during the testing phase for domain adaptation scenarios. AMCD adapts the target features by aligning the predicted outputs of two different classifiers. DSBN allocates a BN layer to each domain to compensate for inter-domain distribution differences. For fair comparison, these comparison algorithms all use UNet as the backbone network.

\begin{figure}[t!]
\centering
    \subfigure[$DRIVE \to  STARE$]{\includegraphics[height=.19\textwidth]{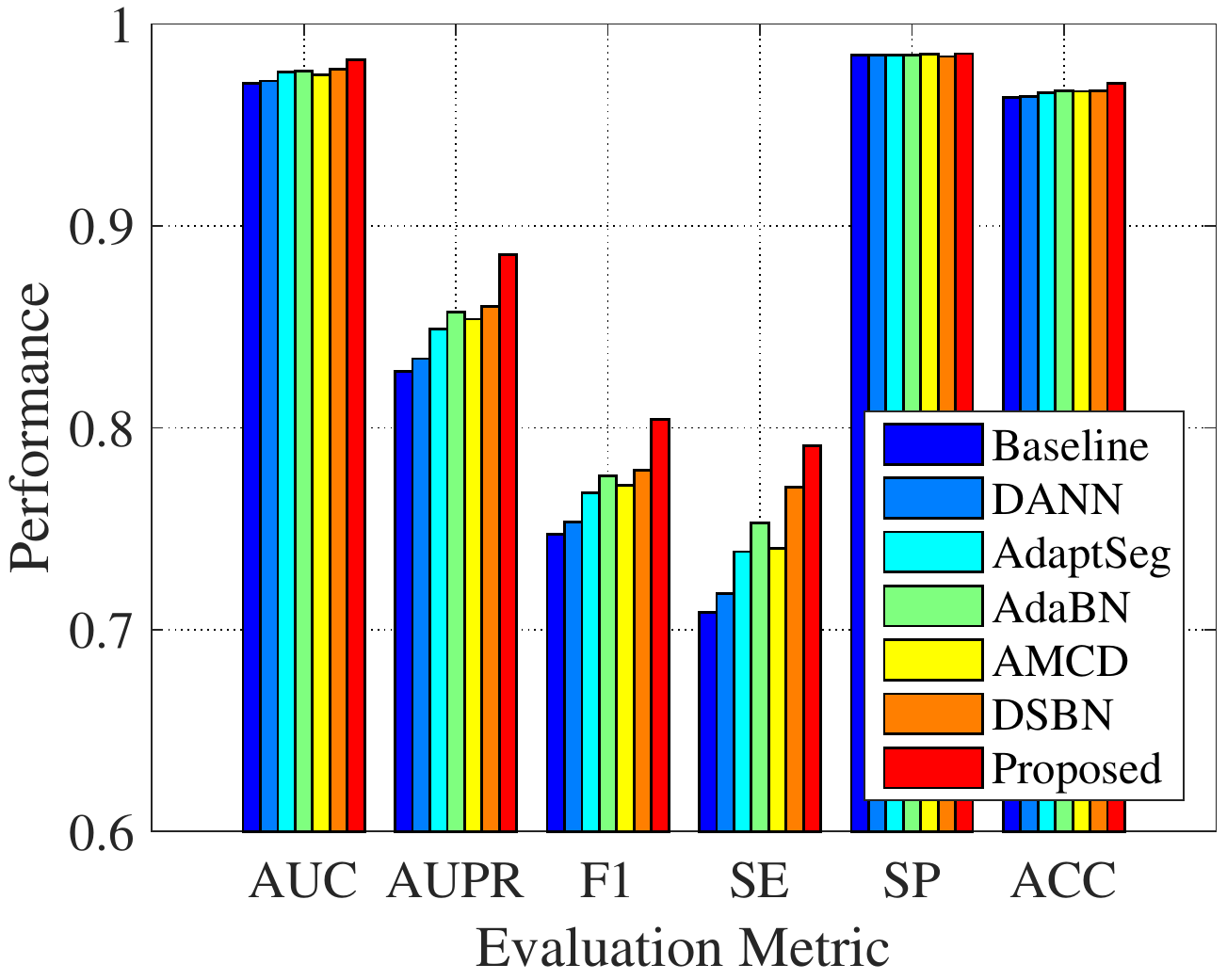}}
    \subfigure[$DRIVE \to  Retinal40$]{\includegraphics[height=.19\textwidth]{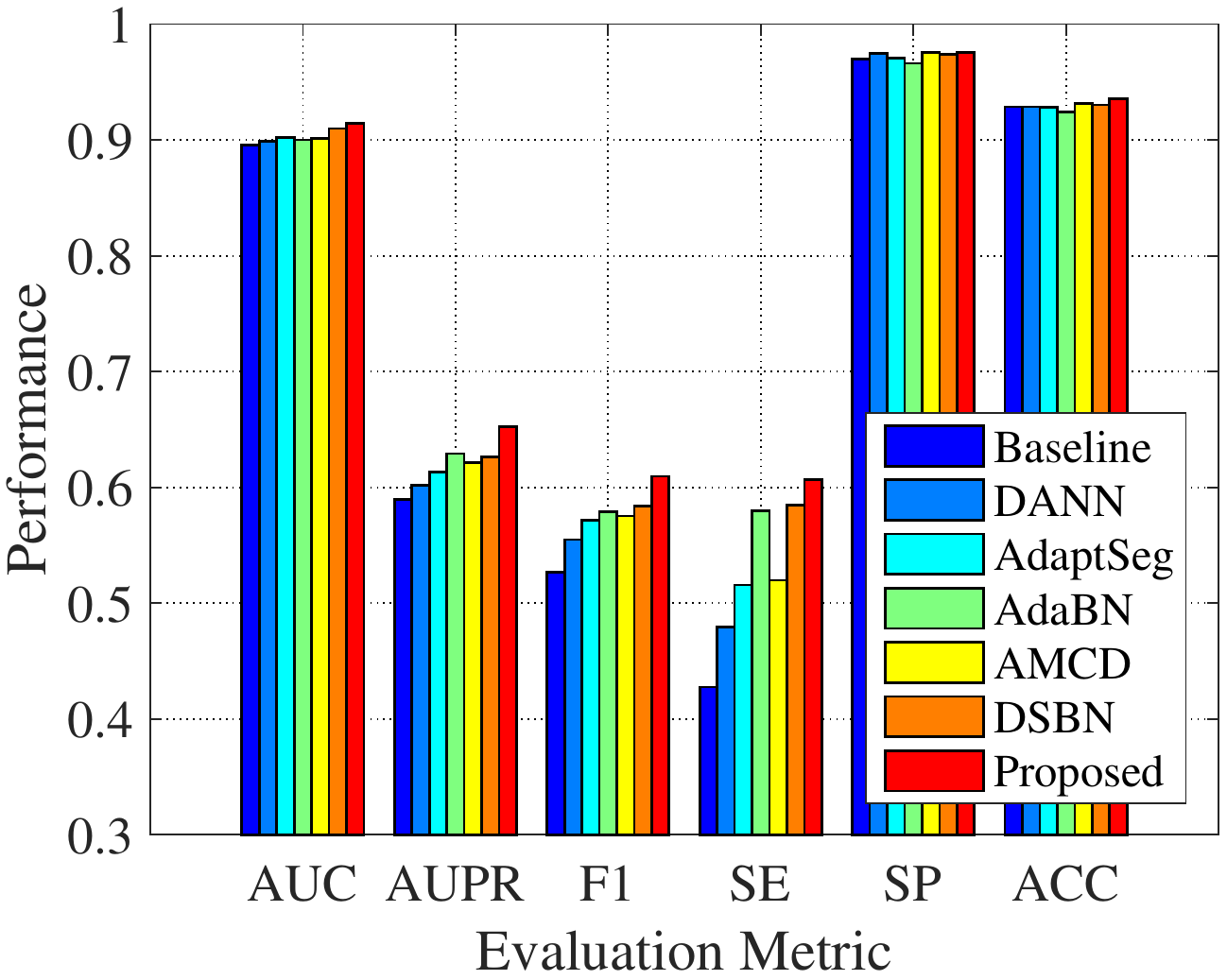}}
    \\
     \vspace{-.1in}
    \subfigure[$STARE \to  DRIVE$]{\includegraphics[height=.19\textwidth]{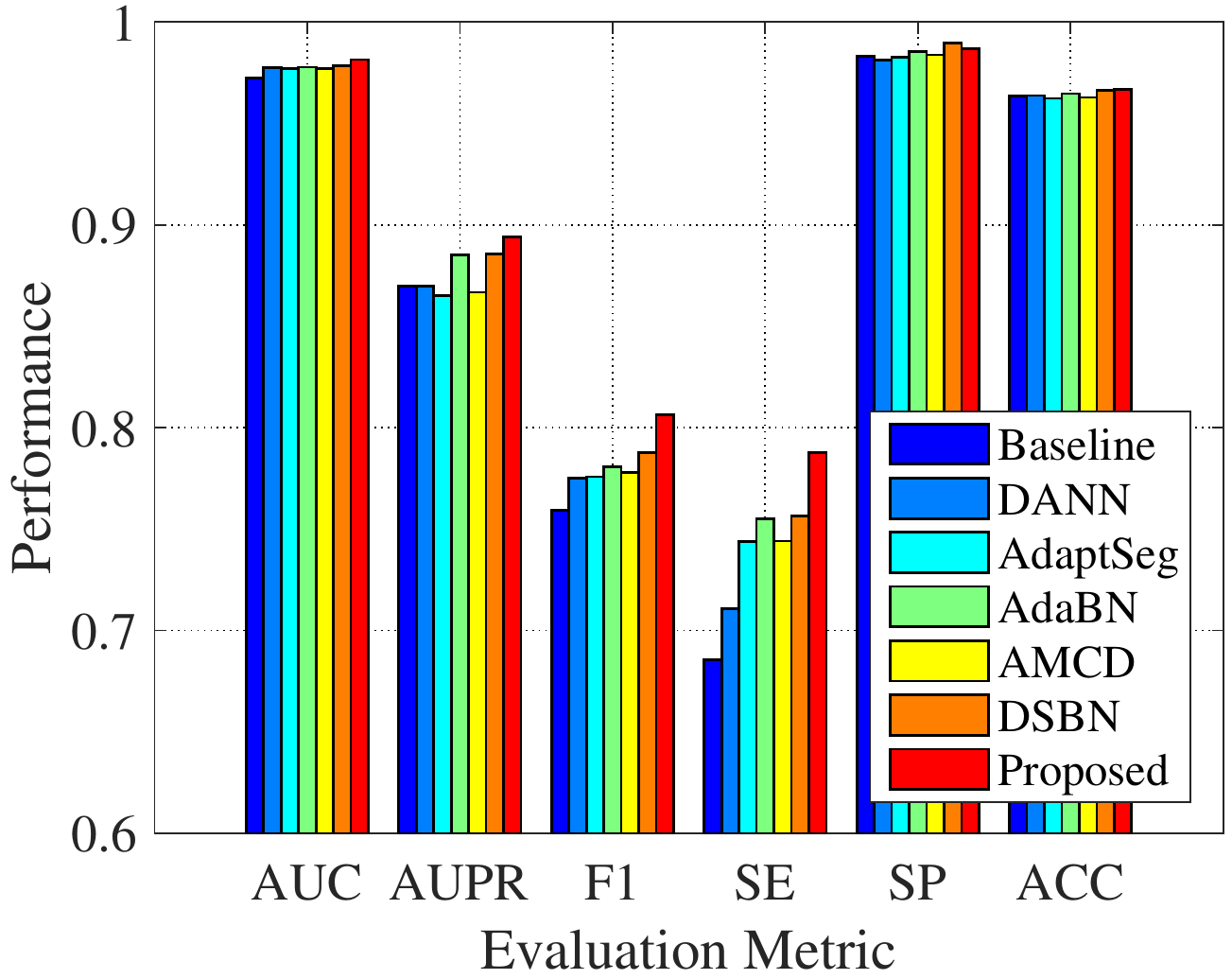}}
    \subfigure[$STARE \to  Retinal40$]{\includegraphics[height=.19\textwidth]{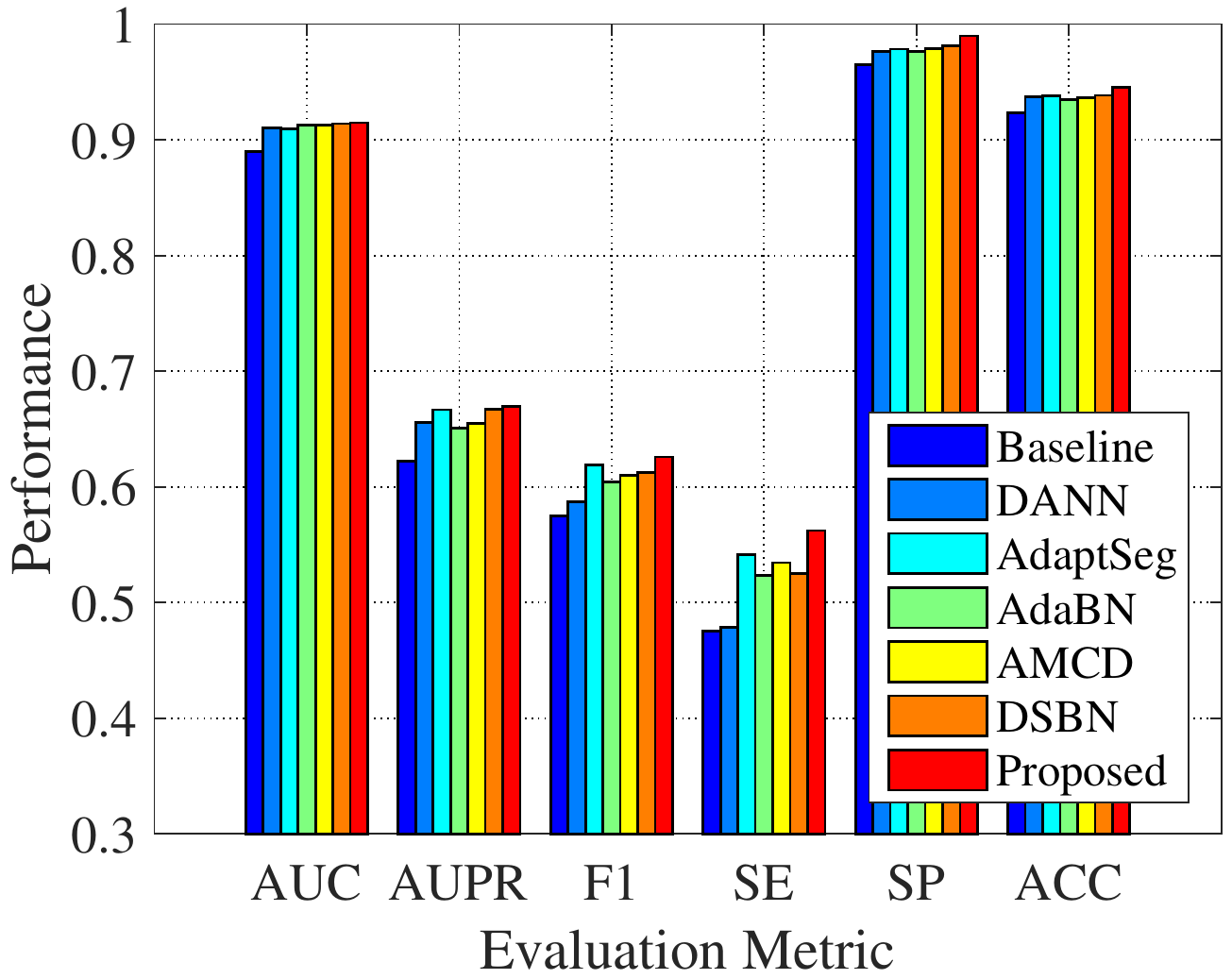}}
    \\
    \vspace{-.1in}
    \subfigure[$Retinal40 \to  DRIVE$]{\includegraphics[height=.19\textwidth]{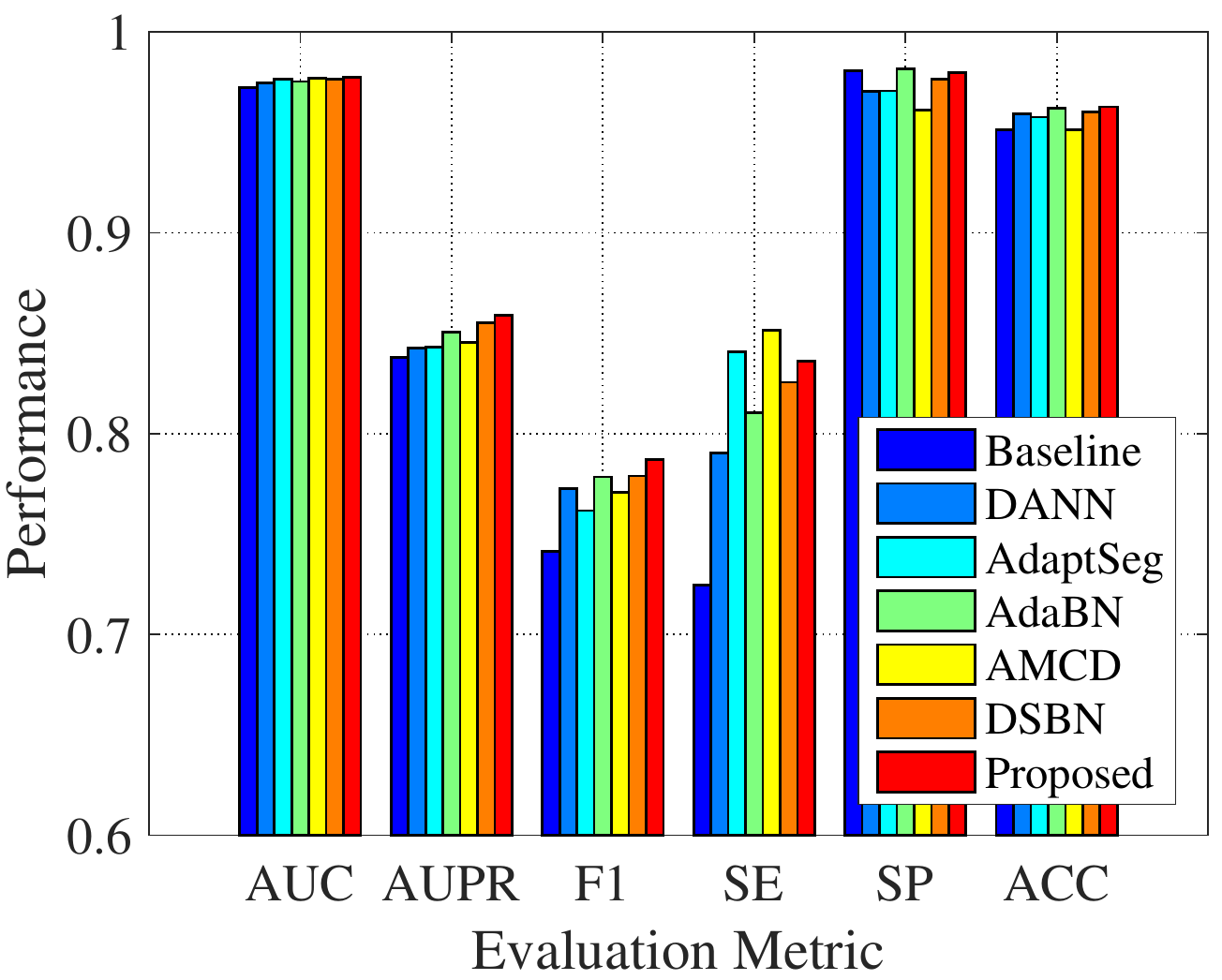}}
    \subfigure[$Retinal40 \to  STARE$]{\includegraphics[height=.19\textwidth]{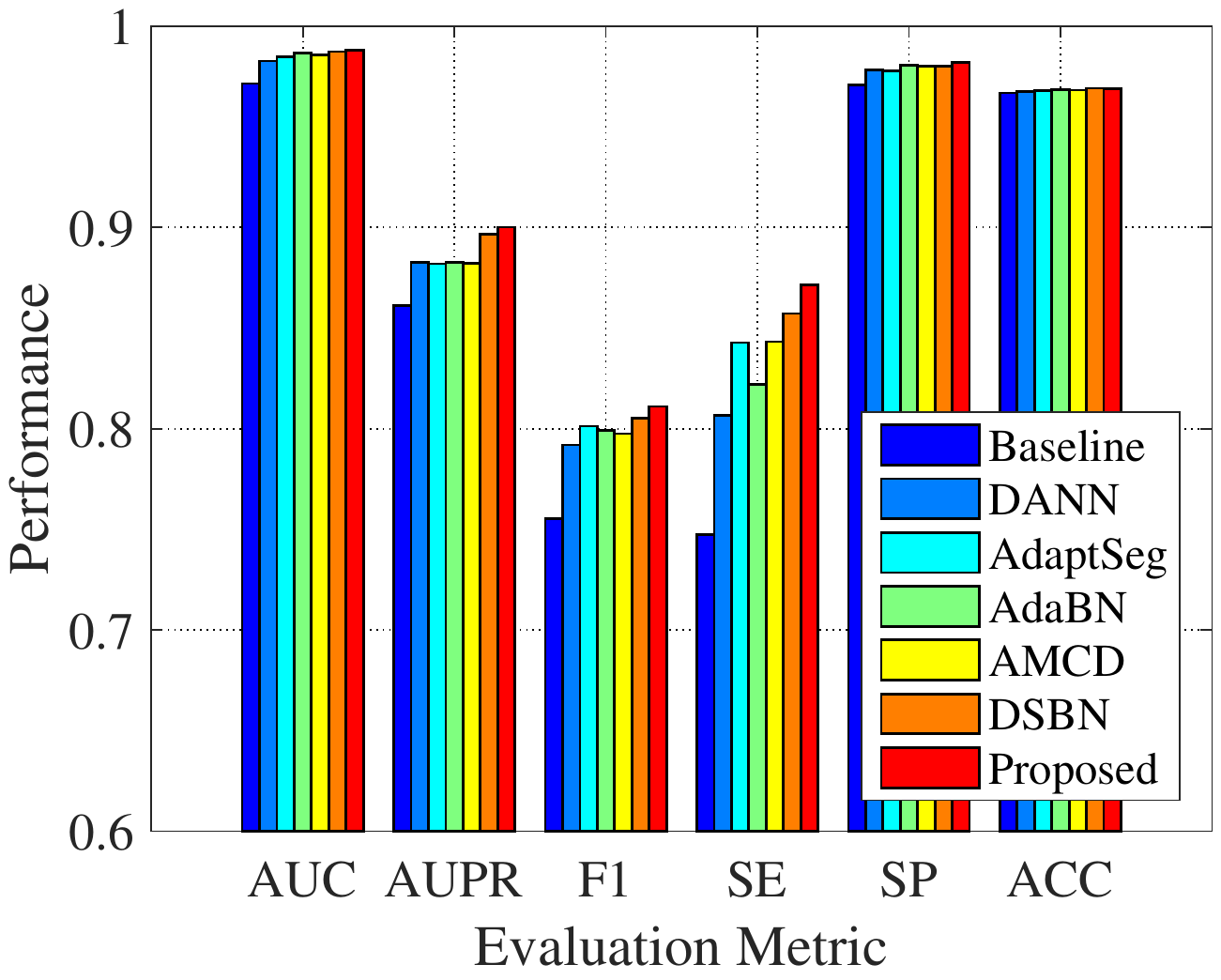}}
    \\
    \vspace{-.1in}
    \caption{Segmentation performance of different algorithms on transfer tasks between regular fundus datasets.}
    \label{Fig.regular}
%   \BSpace
\end{figure}

\figref{Fig.regular} presents the segmentation performance of different algorithms on six
transfer tasks between regular fundus datasets. It can be observed that the proposed method
outperforms the baseline method in all metrics, which demonstrates the need for domain adaptation in the presence of large
differences between domains. In addition, our proposed method achieves better segmentation performance compared to
other adversarial learning methods (DANN,AdaptSeg,AMCD), suggesting that the traditional BN layer may not be
applicable to the domain adaptation problem and that directly combining features from the source and target domains
for normalization deteriorates the transferability of the network. Finally, the proposed method achieves further
performance improvement over AdaBN and DSBN, which indicates that each feature channel may capture different
patterns and our method is able to further improve the generalization performance of the network by reweighting
each feature channel.

\begin{figure*}[t!]
\centering
    % \begin{tabular*}{0.986\linewidth}{|c|c|}
    %     \hline
    %       & \textbf{Channel} $\#23$ ${{\eta }^{\left( 23 \right)}}=0.89$  & \textbf{Channel} $\#16$ ${{\eta }^{\left( 16 \right)}}=0.26$ \\
    %     \hline
    %     \begin{sideways} (a)  $DRIVE \to  UWF$ \end{sideways} &
    %     \includegraphics[width=0.45\linewidth]{ft/img_s_15_channel_23.pdf} &
    %     \includegraphics[width=0.45\linewidth]{ft/img_s_15_channel_16.pdf} \\
    %     \hline
    %     \begin{sideways} (b) $STARE \to  UWF$  \end{sideways} &
    %     \includegraphics[width=0.45\linewidth]{ft/img_t_5_channel_23.pdf} &
    %     \includegraphics[width=0.45\linewidth]{ft/img_t_5_channel_2.pdf}
    %     \\
    %     \hline
    % \end{tabular*},grid,tics=10
    
    \begin{overpic}[width=0.96\linewidth]{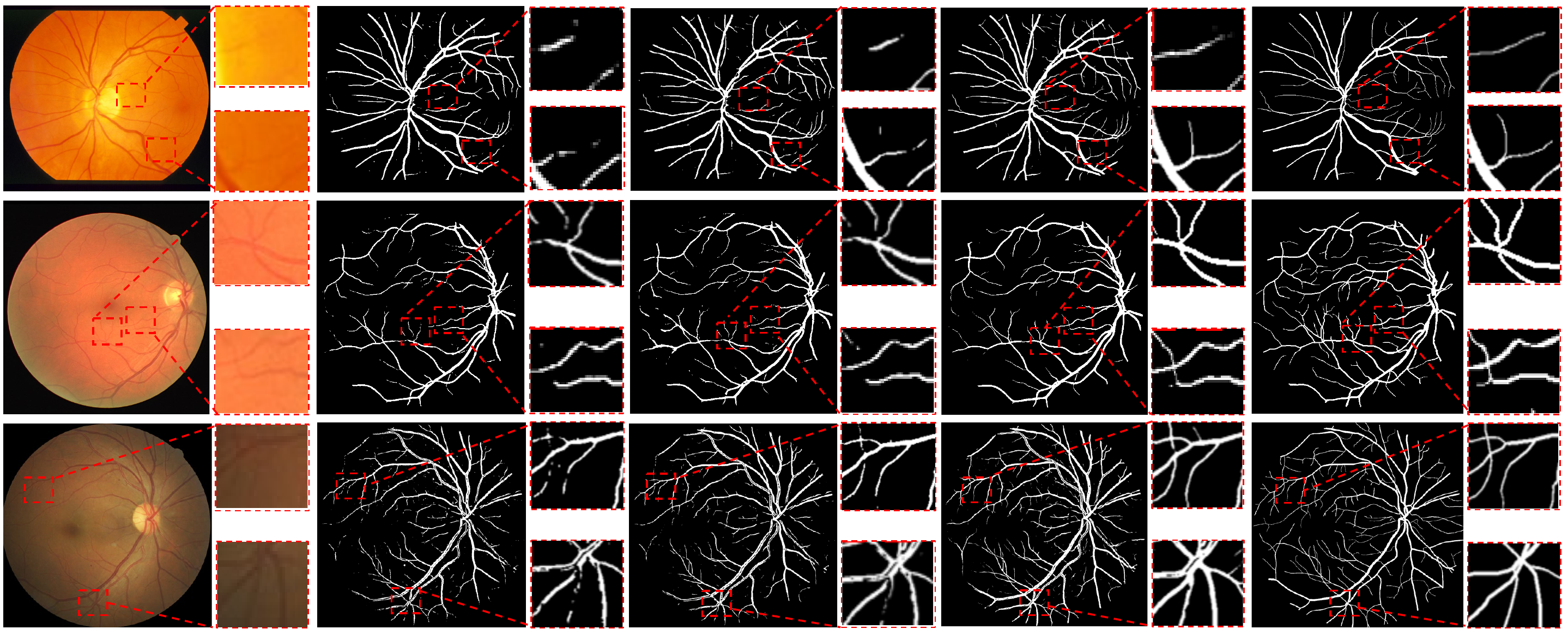}
        \put(9,-2){(a)}
        \put(29,-2){(b)}
        \put(49,-2){(c)}
        \put(69,-2){(d)}
        \put(89,-2){(e)}
    \end{overpic} \\
    \vspace{.10in}
    % \begin{overpic}[width=\linewidth]{seg_result/seg_exp1_2.pdf}
    %     \put(9,-2){(a)}
    %     \put(29,-2){(b)}
    %     \put(49,-2){(c)}
    %     \put(69,-2){(d)}
    %     \put(89,-2){(e)}
    % \end{overpic} \\
    % \vspace{-.10in}
    \caption{Visualization of segmentation results of different algorithms on the transfer tasks between regular fundus datasets. (a) Original image; (b) Baseline; (c) DSBN \cite{liu2020ms}; (d) Proposed; (e) Ground Truth. From top to bottom are the transfer tasks $DRIVE \to  STARE$, $STARE \to  DRIVE$, $DRIVE \to  Retinal40$, respectively. Two enlarged fundus image patches are shown on the right of each subfigure.
    }
    \label{fig.seg_exp1_1}
\end{figure*}
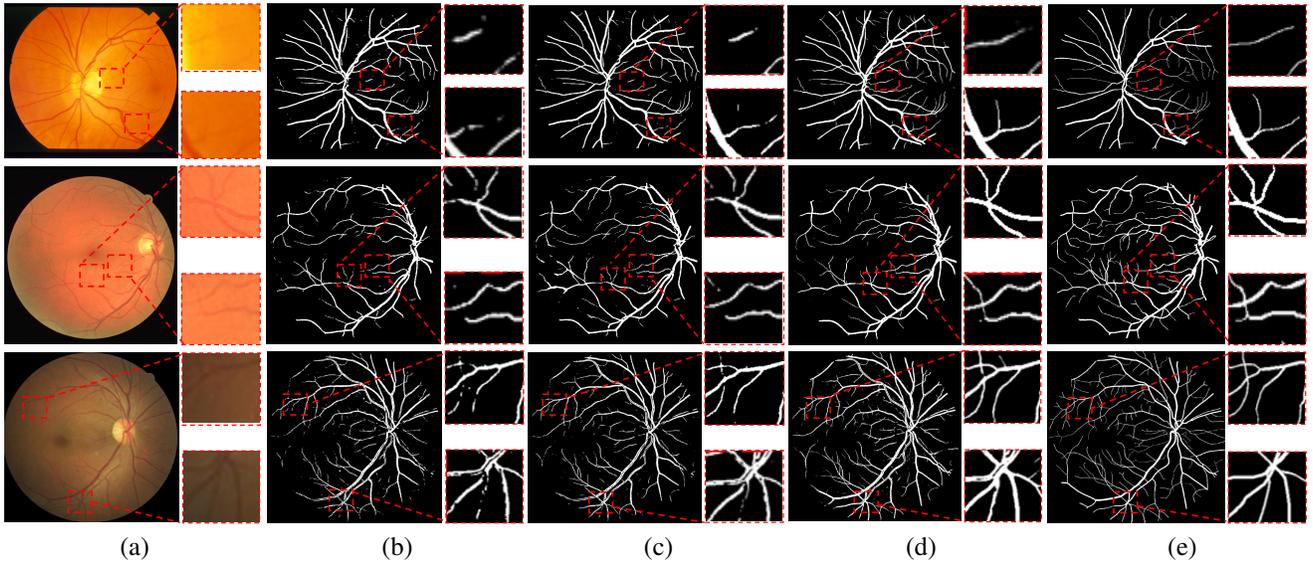

\figref{fig.seg_exp1_1} visualizes the segmentation results of different algorithms on transfer tasks between regular fundus datasets. From top to bottom are the segmentation results for the transfer tasks $DRIVE \to  STARE$, $STARE \to  DRIVE$, $DRIVE \to  Retinal40$, respectively, and the local patches of retinal vessels with different vessel widths are enlarged and shown on the right of each subfigure. It can be found that the proposed method can identify most of the retinal vessels, including those in low contrast and low light conditions. In addition, the proposed method is able to maintain the morphological structure of the vessels well, including the branches of the vessels, the vessel tree, and the edges of the vessels, etc.

% \subsubsection{Segmentation Performance of Transfer Tasks between regular fundus Datasets and UWF Fundus Datasets} \label{sec:exp_2}
\subsubsection{Evaluations on the Transfer Tasks Regular $\to$ UWF}\label{sec:exp_2}
\begin{table}[htbp!]
    \caption{Segmentation performance of different algorithms on the transfer tasks between regular fundus Datasets and UWF Fundus Datasets}
     \label{tab:uwf}
\centering
\setlength{\tabcolsep}{1.5mm}{
\begin{tabular}{cccccccc}
\hline
Method   & \multicolumn{1}{l}{AUC} & \multicolumn{1}{l}{AUPR} & \multicolumn{1}{l}{F1} & \multicolumn{1}{l}{SE} & \multicolumn{1}{l}{SP} & \multicolumn{1}{l}{ACC}  \\
\hline
Baseline & 0.9600                    & 0.6883                 & 0.5224                 & 0.3606                 & 0.9934                 & 0.9836                   \\
% \hline
DANN\cite{javanmardi2018domain}     & 0.9625                  & 0.6950                  & 0.5550                  & 0.4149                 & 0.9953                 & 0.9860                    \\
% \hline
AdaptSeg\cite{tsai2018learning} & 0.9667                  & 0.6965                 & 0.5785                 & 0.4385                 & 0.9962                 & 0.9861                   \\
% \hline
% Minent\cite{grandvalet2005semi}   & 0.9453                  & 0.6483                 & 0.5692                 & 0.4158                 & 0.9992                 & 0.9865                   \\
% \hline
AdaBN\cite{li2016revisiting}    & 0.9720                   & 0.7301                 & 0.6588                 & 0.6091                 & 0.9939                 & 0.9853                   \\
% \hline
AMCD\cite{zhuang2019domain}      & 0.9706                  & 0.7015                 & 0.5973                 & 0.4581                 & 0.9992                 & 0.9865                   \\
% \hline
DSBN\cite{liu2020ms}     & 0.9722                  & 0.7523                 & 0.6974                 & 0.6168                 & \textbf{0.9993}                 & 0.9863                                \\
% \hline
\textbf{Proposed} & \textbf{0.9778}                 & \textbf{0.7707}                 &\textbf{0.7143}                 & \textbf{0.6375}                 & 0.9989                & \textbf{0.9883}  \\
\hline
\hline
Baseline & 0.9601                  & 0.7190                 & 0.6449                 & 0.5301                 & 0.9883                 & 0.9769                   \\
% \hline
DANN\cite{javanmardi2018domain}     & 0.9747                  & 0.7234                 & 0.6628                 & 0.5429                 & 0.9975                 & 0.9874                   \\
% \hline
AdaptSeg\cite{tsai2018learning} & 0.9759                  & 0.7336                 & 0.6745                 & 0.5578                 & 0.9977                 & 0.9875                   \\
% \hline
% Minent\cite{grandvalet2005semi}   & 0.9738                  & 0.7165                 & 0.6535                 & 0.5379                 & 0.9948                 & 0.9865                   \\
% \hline
AdaBN\cite{li2016revisiting}    & 0.9781                  & 0.7561                 & 0.6841                 & 0.5774                 & \textbf{0.9979}                 & 0.9877                   \\
% \hline
AMCD\cite{zhuang2019domain}      & 0.9780                   & 0.7451                 & 0.6814                 & 0.5616                 & 0.9973                 & 0.9872                   \\
% \hline
DSBN\cite{liu2020ms}     & 0.9784                  & 0.7628                 & 0.6856                 & 0.5898                 & 0.9978                 & 0.9879                   \\
% \hline
\textbf{Proposed} & \textbf{0.9811}                  & \textbf{0.7810}                  & \textbf{0.7107}                & \textbf{0.6209}                 & 0.9953                 & \textbf{0.9881}  \\
\hline
\hline
Baseline & 0.9701                  & 0.7568                 & 0.6219                 & 0.4623                 & 0.9894                 & 0.9821                   \\
% \hline
DANN\cite{javanmardi2018domain}     & 0.9768                  & 0.7616                 & 0.6503                 & 0.5442                 & 0.9939                 & 0.9836                   \\
% \hline
AdaptSeg\cite{tsai2018learning} & 0.9776                  & 0.7742                 & 0.6596                 & 0.5936                 & 0.9941                 & 0.9850                    \\
% \hline
% Minent\cite{grandvalet2005semi}   & 0.9759                  & 0.7112                 & 0.6481                 & 0.5302                 & 0.9906                 & 0.9801                   \\
% \hline
AdaBN\cite{li2016revisiting}    & 0.9787                  & 0.7843                 & 0.6685                 & 0.6099                 & 0.9954                 & 0.9866                   \\
% \hline
AMCD\cite{zhuang2019domain}      & 0.9785                  & 0.7765                 & 0.6672                 & 0.6017                 & 0.9949                 & 0.9850                    \\
% \hline
DSBN\cite{liu2020ms}     & 0.9788                  & 0.7830                  & 0.7028                 & 0.6258                 & 0.9959                 & 0.9867                   \\
% \hline
\textbf{Proposed} & \textbf{0.9869}                &\textbf{0.8088}               & \textbf{0.7309}                 & \textbf{0.6634}               &\textbf{0.9964}                & \textbf{0.9888}     \\
\hline
\end{tabular}}
\end{table}
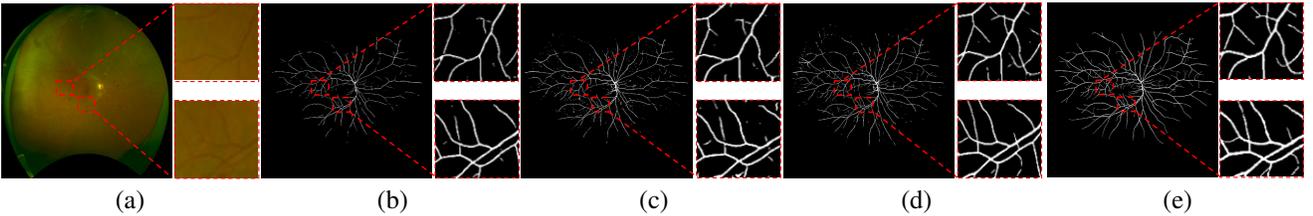
\begin{figure*}[t!]
    \centering
    \begin{overpic}[width=0.96\linewidth]{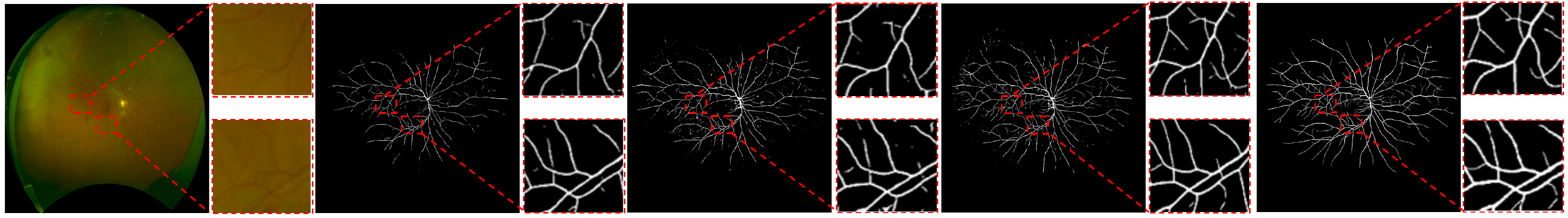}
        \put(9,-2){(a)}
        \put(29,-2){(b)}
        \put(49,-2){(c)}
        \put(69,-2){(d)}
        \put(89,-2){(e)}
    \end{overpic}
    \vspace{.10in}
        % \TSpace
    \caption{Visualization of segmentation results of different algorithms on the transfer task $Retinal40 \to  UWF$. (a) Original image; (b) Baseline; (c) DSBN \cite{liu2020ms}; (d) Proposed; (e) Ground Truth. Two enlarged fundus image patches are shown on the right of each subfigure.
    }\label{fig:uwfvis}
    \BSpace
\end{figure*}
To further validate the effectiveness of the proposed algorithm, we also conducted experiments on the transfer task between the regular fundus dataset and the UWF fundus dataset, which is a more challenging task because along with contrast, lighting conditions, and lesions, the UWF images also have a larger field of view and higher resolution compared to the regular fundus images. The segmentation performance of different algorithms on three transfer tasks $DRIVE \to  UWF$, $STARE \to  UWF$, and $Retinal40 \to  UWF$ 
are listed in the second row, third row and fourth row of \tabref{tab:uwf}, respectively. It can be found that the proposed method outperforms other six comparison methods in almost all metrics on the three transfer tasks. Our method achieves the highest average AUC of 0.9819, the highest average AUPR of 0.7868, the highest average F1 of 0.7186, the highest sensitivity of 0.6406, and the highest accuracy of 0.9884, which yield performance improvements of 
0.55\%, 2.08\%, 2.34\%, 2.98\%, and 0.14\%, respectively, compared to the best comparison algorithm DSBN. This suggests that despite the greater variations between UWF fundus images and regular fundus images, our method still has better discriminatory ability for vessels and is able to detect more vessels than other methods.

\figref{fig:uwfvis} visualizes the segmentation results of different algorithms on the transfer task $Retinal40 \to  UWF$. Two enlarged local fundus regions are shown on the right of each subfigure. It can be observed that segmentation models trained on regular fundus data have difficulty detecting blood vessels in the posterior pole of the retina due to differences in imaging range; DSBN is able to detect more blood vessels in the posterior pole, because it uses domain-specific normalization layer to compensate for domain gaps. Our method is able to select the more transferable feature channels by reweighting each feature channel, thus allowing better detection of vessels in the retina while maintaining the morphological structure of the vessels in the UWF fundus images.
\subsection{Ablation Study}
Our approach consists of two key components: \textbf{entropy-based adversarial learning} and
\textbf{transfer normalization}. Here we conduct ablation studies to verify the effectiveness
of each component. We compare our approach with (1) the baseline model; (2) using only
entropy-based adversarial learning (EA in \tabref{tab.expab}); and (3) using only transfer
normalization (TN in \tabref{tab.expab}). As shown in \tabref{tab.expab}, both entropy-based
adversarial learning and transfer normalization can bring performance improvements for
cross-domain retinal vessel segmentation. Compared with the baseline model, the AUPR of EA and
TN improved by 1.8\% and 2.99\%, respectively. It is worth noting that our proposed method
achieves better performance by combining these two key components.
\vspace{-.15in}
\begin{table}[htbp!]
\caption{Ablation study on different components}
\label{tab.expab}
\centering
\setlength{\tabcolsep}{1.5mm}{
\begin{tabular}{ccccccc}
\hline
Method   & \multicolumn{1}{l}{AUC} & \multicolumn{1}{l}{AUPR} & \multicolumn{1}{l}{F1} & \multicolumn{1}{l}{SE} & \multicolumn{1}{l}{SP} & \multicolumn{1}{l}{ACC}  \\
\hline
Baseline & 0.9701                  & 0.7568                   & 0.6219                 & 0.4623                 & 0.9894                 & 0.9821                   \\
% \hline
EA       & 0.9779                  & 0.7748                   & 0.6636                 & 0.6036                 & 0.9901                 & 0.9850                    \\
% \hline
TN       & 0.9792                  & 0.7867                   & 0.7122                 & 0.6303                 & \textbf{0.9969}                 & 0.9876                   \\
% \hline
\textbf{Proposed} & \textbf{0.9869}                &\textbf{0.8088}               & \textbf{0.7309}                 & \textbf{0.6634}               &0.9964                & \textbf{0.9888}     \\
\hline
\end{tabular}}
\end{table}
\subsection{Channel Visualization}
\begin{figure*}[htbp!]
    \centering
    \begin{tabular*}{0.99\linewidth}{|c|c|c|}
        \hline
          & \textbf{Channel} $\#23$ ${{\eta }_{23}}=0.89$  & \textbf{Channel} $\#16$ ${{\eta }_{16}}=0.26$ \\
        \hline
        \begin{sideways} Source Domain \end{sideways} &
        \includegraphics[width=0.45\linewidth]{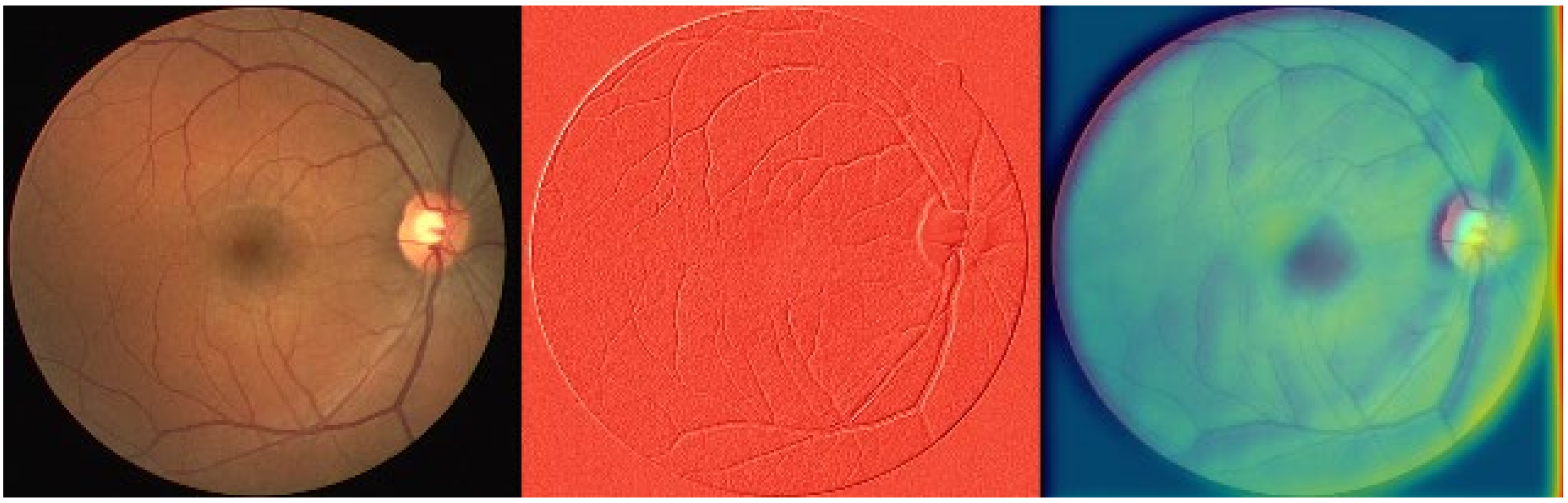} &
        \includegraphics[width=0.45\linewidth]{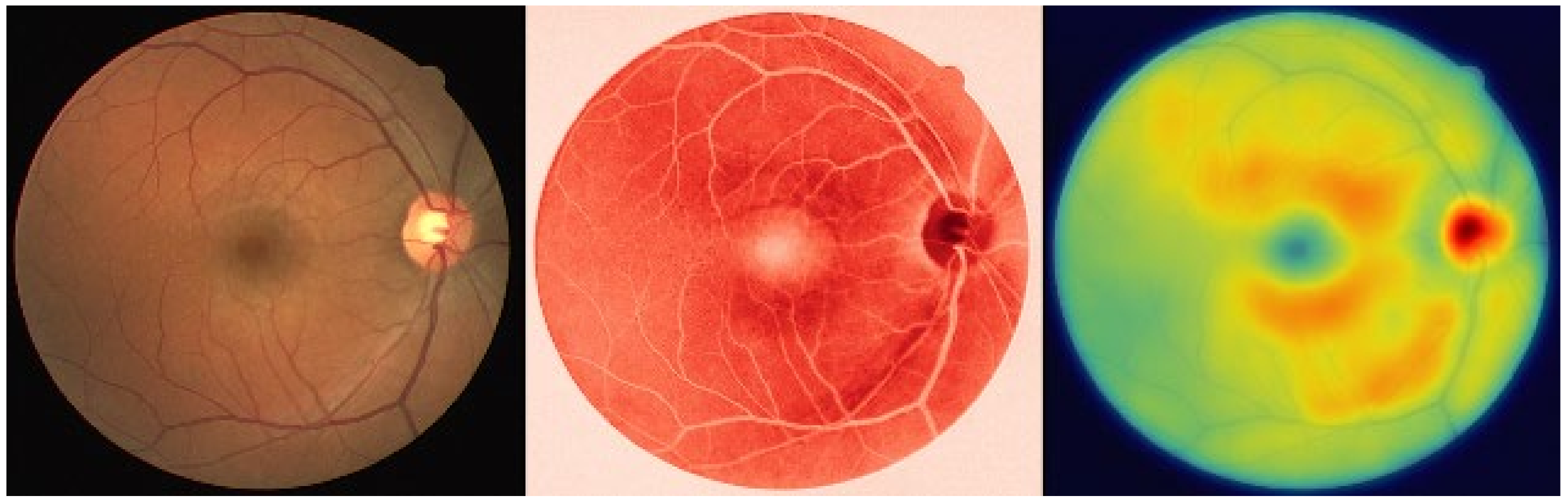} \\
        \hline
        \begin{sideways} Target Domain  \end{sideways} &
        \includegraphics[width=0.45\linewidth]{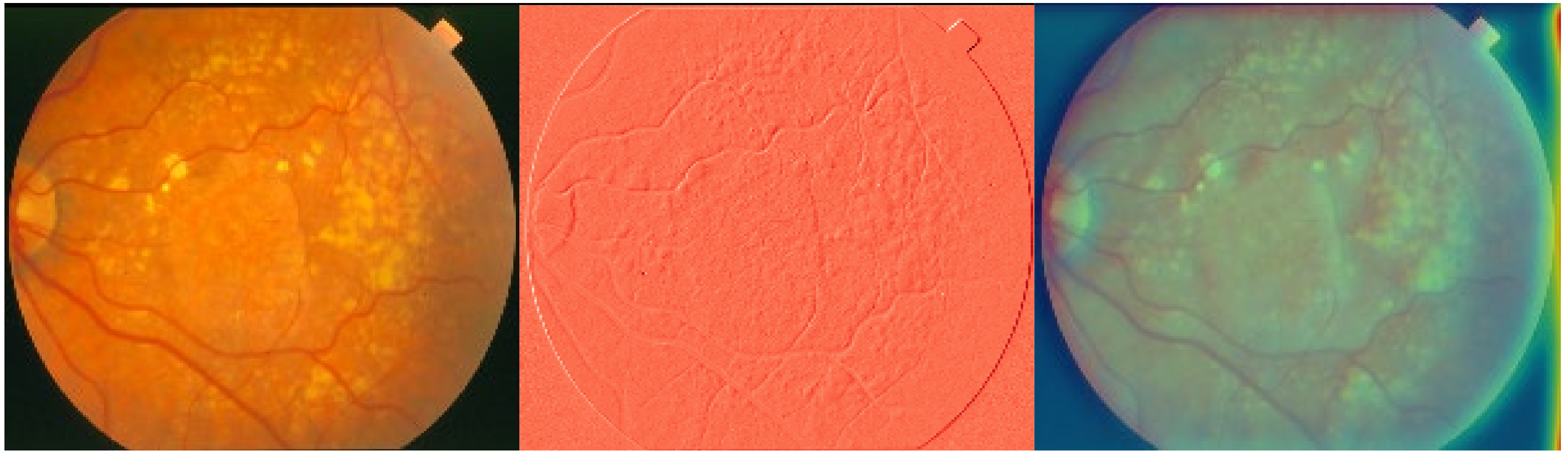} &
        \includegraphics[width=0.45\linewidth]{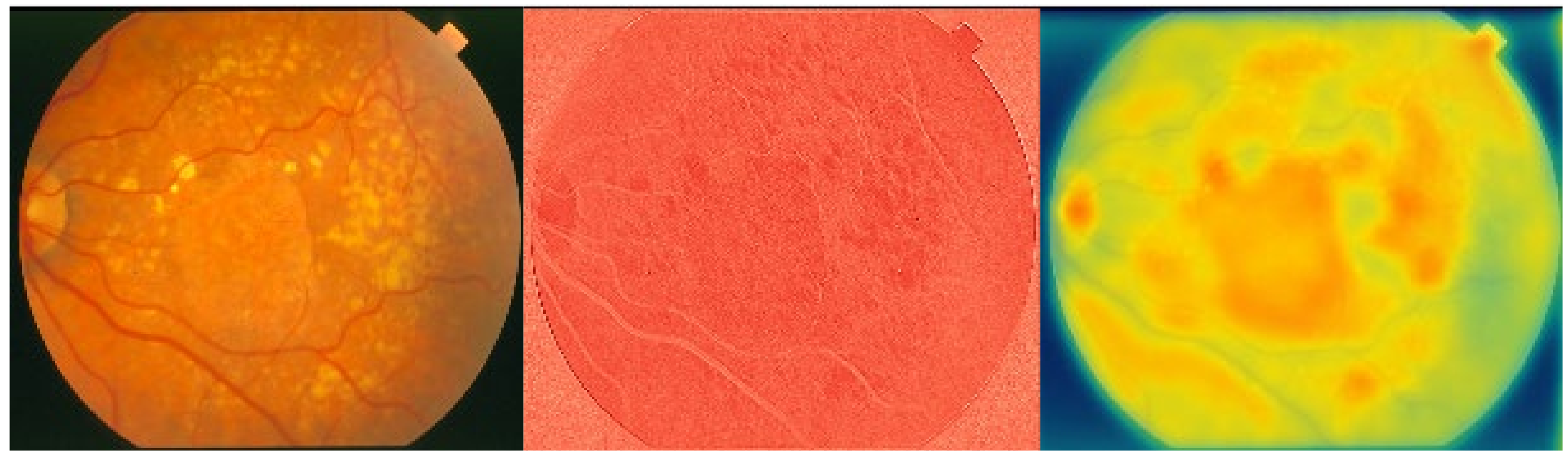}
        \\
        \hline
    \end{tabular*}
    \caption{Visualization of the transferability of different feature channels. In each group, the left is the original image, the middle is the feature channel, and the right is the result of combining the feature channel and the original image.
    }\label{fig:channel}
    \BSpace
\end{figure*}

The weight $\eta$ calculated in \equref{equ:tn_3} quantify the
transferability of the different feature channels, and a
larger weight means that the feature channel is more
transferable and therefore is given higher importance. Here,
we visualize some feature channels. As shown in \figref{fig:channel}, it can be observed that certain feature channels are given a larger weight, e.g., $\eta$ of 0.89 for channel 23, due to the fact that this feature channel captures domain-shared
features, e.g., the texture of blood vessels; while certain
channels are given a smaller weight, e.g., $\eta$ of 0.26 for
channel 16, due to the fact that these feature channels
capture non-essential domain-specific patterns, e.g., backgrounds, lesions,
etc.
\subsection{Discussion}
\subsubsection{Different Distance Type and Different Probability Type}
In our transfer normalization, in order to quantify the transferability of the different feature channels, we first calculate the distances between the different feature channels by \equref{equ:tn_2}, and then convert the calculated distances into probability values by \equref{equ:tn_3} to re-weight each feature channel. Here, we consider different types of distances: mean-based distance, Wasserstein Distance ($||E[h_{m}^{s}]-E[h_{m}^{t}]||_{2}^{2}+((\operatorname{Var}[h_{m}^{s}])^{2}+(\operatorname{Var}[h_{m}^{t}])^{2}-2\operatorname{Var}[h_{m}^{s}]\operatorname{Var}[h_{m}^{t}])$, and our method. We conducted experiments on the transfer task $Retinal40 \to  UWF$. As shown in 
\tabref{tab.expdis}, the mean-based distance does not take into account the variance, while the Wasserstein Distance cannot balance the importance of the mean and the variance. Our method uses the mean normalized by variance, which is similar to BN, and yields the best performance.
In addition, we compared three probability types: Softmax, Gaussian, and Student-t distribution. As shown in \tabref{tab.expdis}, better generalization performance was achieved using the Student-t distribution because it has a heavier tail and therefore it is able to emphasize the more transferable feature channels while avoiding over-punishing the other channels.
% \vspace{-.15in}
\begin{table}[htbp!]
\caption{Segmentation Performance of our method using different distance types and different probability types}
\label{tab.expdis}
\centering
\setlength{\tabcolsep}{0.8mm}{
\begin{tabular}{ccccccc}
\hline
\multirow{2}{*}{Method} & \multicolumn{3}{l}{Different
  distance types} & \multicolumn{3}{l}{Different
  probability types}  
  \\
  \cmidrule(r){2-4} \cmidrule(r){5-7}
    % \cline{2-4}  \cline{5-7} 
                        & ${E[h_{m}]}$      & \makecell[c]{Wasserstein \\ Distance}      & $\frac{E[h_{m}]}{\sqrt{\operatorname{Var}[h_{m}]+\epsilon}}  $                          & Softmax & Gaussian & Student-t                    \\ 
                        \hline
% \cline{1-1}
AUC                     & 0.9822 & 0.9855 & \textbf{0.9869}                       & 0.9834  & 0.9859    & \textbf{0.9869}                       \\
% \hline
AUPR                    & 0.7833 & 0.7953 & \textbf{0.8088}                       & 0.7918  & 0.7985    & \textbf{0.8088}                       \\
% \hline
F1                      & 0.7123 & 0.7156 & \textbf{0.7309}                       & 0.7192  & 0.7241    & \textbf{0.7309}                       \\
% \hline
SE                      & 0.6421 & 0.6542 & \textbf{0.6634}                       & 0.6572  & 0.6592    & \textbf{0.6634}                       \\
% \hline
SP                      & 0.9945 & \textbf{0.9971} & 0.9964                       & \textbf{0.9964}  &  0.9953   & \textbf{0.9964}                       \\
% \hline
ACC                     & 0.9821 & 0.9829 & \textbf{0.9888}                       & 0.9863  & 0.9872    & \textbf{0.9888}     \\
\hline
\end{tabular}}
\end{table}

\subsubsection{Connection to Entropy Minimization Principle}
\begin{figure}[htbp!]
    \centering
    \begin{overpic}[width=\linewidth]{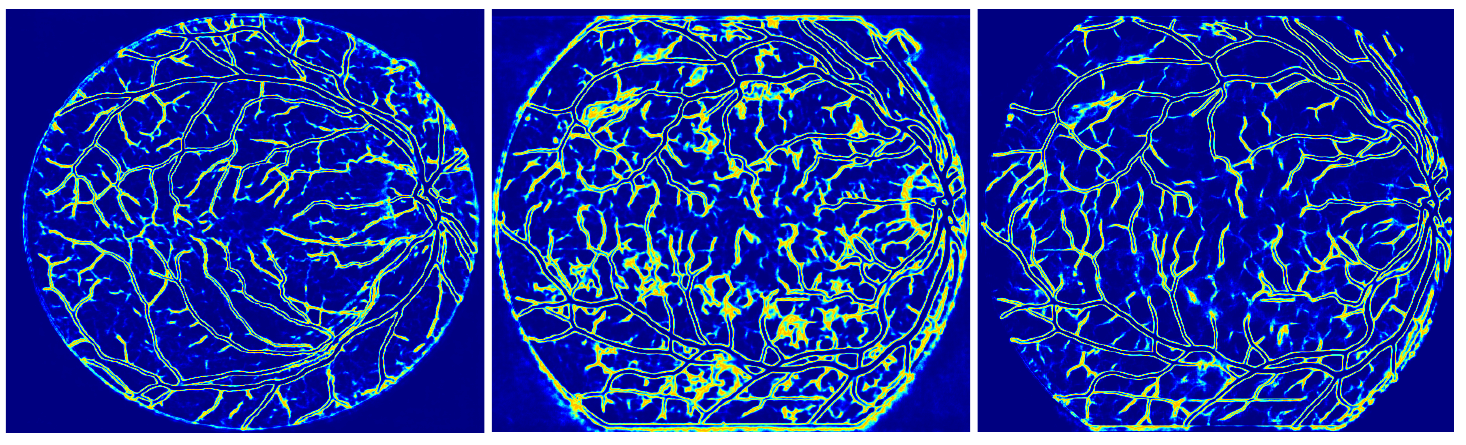}
        \put(15,-3){(a)}
        \put(49,-3){(b)}
        \put(82,-3){(c)}
        % \put(69,-2){(d)}
        % \put(89,-2){(e)}
    \end{overpic} \\
        \vspace{.10in}
    \caption{Visualization of entropy maps. (a) entropy map on the source domain fundus dataset; (b) entropy map produced by the baseline method on the target domain fundus image; (c) entropy map produced by our method on the target domain fundus image.}
    \label{fig:ent}
        % \BSpace
\end{figure}
\cite{long2018transferable} proposed to apply the entropy minimization principle to the domain adaptation setting to adjust the model to pass through low-density region of unlabeled target domain data, so that the model can make highly confident predictions on target domain data. The entropy-based adversarial learning in our approach can be seen as an indirect entropy minimization. As shown in \figref{fig:ent}(a),(b), due to domain shift, the entropy maps of the source and target domains have a large difference in distribution, where the entropy map of the source domain is clear (certain,low-entropy) while that of the target domain is noisy (uncertain,high-entropy). Our proposed method aligns the distribution of the entropy maps of the source and target domains by using adversarial learning, thus encouraging the model to be able to produce highly confident predictions even on unlabeled target domain data (see \figref{fig:ent}(c)). 
Furthermore, our approach and the entropy minimization principle are complementary. Here, we conducted experiments on the transfer task $Retinal40 \to  UWF$. As shown in \tabref{tab.expent}, incorporating entropy minimization principle (MinEnt) into our proposed method yields better segmentation performance.
\vspace{-.15in}
\begin{table}[htbp!]
\caption{Segmentation performance of our proposed method and our proposed method with MinEnt}
\label{tab.expent}
\centering
\setlength{\tabcolsep}{1.mm}{
\begin{tabular}{ccccccc}
\hline
Method   & \multicolumn{1}{l}{AUC} & \multicolumn{1}{l}{PR} & \multicolumn{1}{l}{F1} & \multicolumn{1}{l}{SE} & \multicolumn{1}{l}{SP} & \multicolumn{1}{l}{ACC}  \\
\hline
Proposed & 0.9869                &0.8088               & 0.7309                & 0.6634               & \textbf{0.9964}                & 0.9888     \\
% \hline
Proposed+MinEnt\cite{long2018transferable} & \textbf{0.9871}                &\textbf{0.8093}               & \textbf{0.7321}                 & \textbf{0.6642}               &\textbf{0.9964}                & \textbf{0.9891}     \\
\hline
\end{tabular}}
\end{table}
\section{Conclusion}
In this paper, we propose a new deep unsupervised domain adaptation framework for cross-domain retinal vessel segmentation. Using our proposed entropy-based adversarial learning and transfer normalization layer, we can train a segmentation model to be able to generalize well across fundus datasets from different domains, without requiring any ground truth labels for fundus images in the target domain. Extensive experiments on regular fundus datasets and UWF fundus dataset validate the effectiveness of the proposed approach. Our approach not only brings a new idea for unsupervised domain adaption in retinal vessel segmentation, but also presents an innovative approach that can potentially be used for other cross-domain medical image segmentation tasks.

\bibliographystyle{IEEEtran}
\bibliography{DA}

\vspace{-.4in}
\end{document}